\documentclass[a4paper]{PoS}
\title{Simulation of Near Horizontal Muons and Muon Bundles for the HAWC Observatory with CORSIKA}

\ShortTitle{Simulation of Near Horizontal Muons}

\author{Ahron S. Barber$^a$ , David B. Kieda$^{a}$  and \speaker{R. Wayne Springer}$^a$  
and for the HAWC Collaboration$^b$ \\
        \llap{$^a$}Department of Physics and Astronomy, 
        University of Utah, Salt Lake City, UT, USA \\
        \llap{$^b$}For a complete author list, see 
        \href{http://www.hawc-observatory.org/collaboration/icrc2017.php}{www.hawc-observatory.org/collaboration/icrc2017.php}.
        Email: \email{ahron.barber@utah.edu}, \email{dave.kieda@utah.edu}, \email{wayne.springer@utah.edu}}

\abstract{The HAWC (High Altitude Water Cerenkov) gamma ray observatory observes muons with nearly-horizontal trajectories corresponding to zenith angles greater than $80^{0}$. HAWC is located at an altitude of 4100 meters a.s.l. (70 deg. atmospheric depth of 2400 g/cm$^{2}$) on the extinct volcano, Sierra Negra in Mexico. In this poster, we summarize the CORSIKA and GEANT4  as well as toy-model based simulations performed to determine the effective area  of HAWC to muons from high zenith angle cosmic ray primaries. We are developing an updated GEANT4 based detector response simulation that includes a model of the volcanoes that are located near HAWC. These simulations are investigating the capability to use muon multiplicity and rates to differentiate between the primary particle composition (proton or iron) and measure the primary energy.}
\FullConference{35th International Cosmic Ray Conference --- ICRC2017\\
		10--20 July, 2017\\
		Bexco, Busan, Korea}

\begin{document}

\section{INTRODUCTION}
This report describes a study of the response of the HAWC observatory \cite{refHAWC} to muons with nearly horizontal trajectories.   The response to these muons is complicated due to the presence of nearby volcanoes as well as instrumental effects. The minimum threshold energy required for muons to traverse the overburden depth of the volcanoes depends upon muon arrival direction. This direction dependent energy threshold  enables a measurement of the energy spectrum of cosmic rays using  observed muon rates as a function of arrival direction.  Understanding these muon rates is also useful in the determination of backgrounds to neutrino searches \cite{refHAWCNEUTRINO} performed with the HAWC.   We are investigating the possibility to measure the cosmic ray primary composition as well as energy spectrum  using muon bundles observed with HAWC. For all of these analyses,  the effective area as a function of arrival direction for the detection of cosmic ray primaries that produce nearly horizontal muons must be determined. The following sections describe the HAWC detector and the simulation tools used to determine its response to nearly horizontal muons. 

\section{HAWC Detector and Site}
 The HAWC observatory detector consists of three hundred 4.5 m high, 7.3 m diameter tanks covering a footprint of roughly 22,000 $m^{2}$. Each tank contains 200,000 liters of purified water and is instrumented with 4 upward looking photomultiplier tubes (PMTs) . The PMTs can resolve the arrival time of Cherenkov light pulses to better than 1 ns with good  charge resolution.  The HAWC configuration of isolated tanks equipped with PMTs with such precise and accurate timing allows the detector to be operated as a muon hodoscope that can identify muons traversing from tank to tank at a particle propagation speed consistent with the speed of light. The charge measurements can be used  to locate the muon trajectory inside of a tank as well as to determine the number of muons traversing a tank to identify muon bundles.  Enhancements of response in certain azimuth directions arise due to the alignment of the tanks as indicated by the layout of the HAWC detector PMTs shown in Figure 1a. 

The HAWC observatory site is located at an altitude of 4100m  a.s.l. adjacent to two volcanoes, Pico de Orizaba and Sierra Negra in the state of Puebla, Mexico. Sierra Negra extends up to $20^{0}$  in elevation into the field of view of HAWC. The broadest distance  through Sierra Negra  at the altitude of HAWC is 7200m. Pico de Orizaba rises to $14^{0}$  in elevation with a broadest distance of 12000m. The corresponding rock overburden depths vary from 0 to approximately 32000 meter water equivalent (m.w.e).   The minimum energy required by a muon to traverse a given amount of material with an exit energy of at least 50 GeV was determined using the information found in the particle data group chapter on cosmic ray muons \cite{refPDG}.  The overburden depth at the base of Sierra Negra of 7200 m.w.e results in a muon energy threshold of approximately 3.3 TeV, while at the base of Pico de Orizaba the overburden depth of 32000 m.w.e  results in a muon energy threshold of approximately 520 TeV.  The minimum required muon energy to traverse the overburden depth due to the volcanoes Pico de Orizaba and Sierra Negra as a function of arrival direction is shown in Figure 2a. For elevation angles below $-2^{0}$, the muon trajectories intersect the Earth resulting in even higher energy thresholds.  The muon rate impinging on HAWC is therefore modulated by arrival direction due to this varying overburden depth and muon energy threshold.  

\paragraph*{Nearly-Horizontal Muon Identification and Reconstruction}
The procedure to identify nearly-horizontal muons traversing the HAWC detector is to identify patterns of PMT hits in time and space that are consistent with a particle moving nearly horizontally at the speed of light. The identification is done by first identifying a set of PMTs whose locations and hit times satisfy the relation $distance =c\Delta t$. Application of a Hough transformation algorithm \cite{refHOUGH} to further require isolated lines reduces the background from air shower events. This simulation study has aided the development and verification of the horizontal muon tagging and reconstruction algorithms. A description of these algorithms and the observation of nearly horizontal muons by the HAWC gamma ray observatory is described in another report that has been submitted to this conference \cite{refNHM}.

%%
%%%%%%%%%%%%%%%%%%%%%%%%%%%%%%%%%%%%%%%%%%%%%%%%%%%%%%%%%%
%%
\section{Simulation of HAWC detector response to horizontal muons}
 We developed tools required to evaluate the response of HAWC to nearly horizontal muon trajectories. To study the response to cosmic ray primaries, such as protons and nuclei, we utilize CORSIKA  (version 7.5 64bit) simulation software package \cite{refCORSIKA} to generate extensive air showers (EAS)  in the atmosphere.  The trajectories and momenta of the shower particles at the observation level from CORSIKA are then passed to either a generic cylindrical "toy model" or detailed GEANT4-based \cite{refGEANT4} detector model simulation. We utilize the "toy model" to perform faster simulations to develop an understanding of the general behavior and response of a perfect detector.   We inject single muon particle trajectories and momenta directly into our detailed detector simulation to study basic instrumental response. The resulting simulated data is then processed with the nearly-horizontal muon identification and reconstruction software.  We account for the effect of volcano overburden by calculating an arrival direction dependent attenuation of the impinging muon flux due to a minimum required muon energy. A comparison between thrown and reconstructed distributions is then performed to determine acceptances and observable resolutions. A full detector simulation study, including a detailed GEANT4-based description of the  volcanoes, has not yet been completed.

\paragraph*{Extensive Air Shower (EAS) simulation with CORSIKA}
These studies require simulating muon production that takes place in EAS initiated at distances up to many hundreds of kilometers from the HAWC detector. Therefore,  unlike the configuration used in standard HAWC simulations, CORSIKA was configured in our studies for a non-flat detector using a curved atmosphere model.
Our simulations begin by using  CORSIKA to generate EAS from primary particles such as protons and iron. Primary particles are generated using QGSJet II + Gheisha hadronic interaction models over energy ranges and arrival directions that have appreciable rates of nearly horizontal muons observable by the HAWC detector.  The primary particle energy ranges considered in this study range from a minimum of 100 GeV to several PeV. The primary particle arrival directions are generated  for  zenith angles, $\theta$,  greater than $80^{0}$ with probability weighted by $sin(\theta)$  and uniformly over all azimuth angles. By definition, the EAS shower axis  always intersects the origin of the observation level coordinate system. Particle trajectories and momenta are recorded at the observation level of 4100 m  altitude in this coordinate system.   The effect of the Earth's magnetic field at the HAWC site  ($B_{x}=27.17\mu T$ , $B_{z}=29.9 \mu T$) on the muons trajectories was considered. The CORSIKA shower data files are then used by our "toy model" and detailed detector simulation programs. 
 
We typically use ("throw") a CORSIKA generated shower from 100 up to 500 times in our simulation programs by displacing the shower axis (and the shower particles at the observation level) in a plane perpendicular to the original shower axis. The displacement is a randomly chosen perpendicular distance from the original shower axis $R_{p}$ along a plane angle $\phi$ to uniformly populate the circular area out to a distance $R_{p,max}$.  $R_{p,max}$  should be such that the displacement area  exceeds the region of efficient detector acceptance.  In a later section, we estimate $R_{p,max}$ using geometrical considerations and the  expected opening angles between the muons and the shower axis as determined by CORSIKA simulations.  We use a "toy model" simulation to determine $R_{p,max}$  for subsequent detailed GEANT4-based simulations. The effective area for the detection of muons generated by EAS of the cosmic ray primaries from a given arrival direction  is determined by multiplying the area of this disk, $\pi R_{p,max}^{2}$ , by the ratio of the number of events with detected muon(s) to "thrown" primary particles.

Simulating air showers and tracking the particles at the observation level is readily performed for zenith angles up to $90^{0}$ using CORSIKA when configured appropriately. Simulating earth skimming air showers whose zenith angles exceed $90^{0}$ yet produce nearly-horizontal muons that impinge upon HAWC is not directly possible using standard configurations of running CORSIKA simulations. It is possible to obtain a zeroth order estimate of acceptance at nearly-horizontal trajectories slightly beyond $90^{0}$ in zenith by using symmetry arguments by stating that the acceptance for the region between $90^{0}$ and $97^{0}$ is equal to the acceptance between $83^{0}$ and $90^{0}$.  We developed, using the toy model simulation, a technique to use air showers and their resulting tracks by mirroring the generation of air showers about zenith angle = $90^{0}$ by rotating the particle trajectories and momenta appropriately.    This technique provides an improved estimate of the acceptance for zenith angles beyond $90^{0}$ but still ignores subtle details of the differences in the early shower development at the larger zenith angles.

\paragraph*{CORSIKA ``Tree-level''  Observable Distributions}
The region of the Earth's atmosphere where muon production  may occur that results in a detected nearly-horizontal muon is vast. Guidance on configuring the simulation parameters can be obtained by considering that the geometry of large zenith angle showers is dictated by the curvature of the earth and the height of the detector above the surrounding landscape.   Further guidance can be obtained by examining basic distributions of observables from CORSIKA air showers.     The distance to the production location of nearly-horizontal muons can reach approximately 1290 km for muon production at the most probable first interaction height of 83km for horizontal showers. This is estimated by the distance from HAWC to the intersection point of a line tangent to the Earth's surface at HAWC intersecting a circle of radius  $R_{earth}+ 83 km$. The time required for a  muon traveling 1290 km from its creation point is approximately 4.3 milliseconds.  This corresponds to a relativistic gamma factor that requires muon energies of at least 200 GeV.  These geometrical calculations are consistent with a distribution of muon flight distances obtained from a CORSIKA simulation for muons detected  with a simplified perfect detector model from showers of zenith angles up to $90^{0}$. 

A sample of extensive air showers initiated by protons with energy between 100 GeV and 5 PeV with a spectral index of -2.7 was used to obtain the distribution of opening angle between the muon direction and shower axis as a function of momentum. It was determined that the maximum opening angle between the muon direction and shower axis for muons with momentum > 100 GeV/c  was $5^{0}$ . The elevation angle of a ray originating from the HAWC site at elevation 4100 m intersecting with the earth's surface at sea level is approximately $-2^{0}$ , corresponding to a maximum unobstructed zenith angle of $92^{0}$. Accounting for a $5^{0}$ opening angle between shower axis and muon trajectory, the maximum zenith angle for a shower that could possibly produce a muon whose trajectory could traverse the atmosphere unobstructed to HAWC would then be $97 ^{0}$. We therefore limit our throwing of air showers to a maximum of $97^{0}$ in zenith angle. For zenith angles above $90^{0}$ we use the the rotation techniques noted above.  For higher momenta the opening angle of muons from the shower axis decreases to be significantly less than a degree.  The tighter collimation of the muon trajectories at higher energies allow bundles of muons to be observed by the HAWC detector. We have seen muon bundle candidates in the real data as well as in CORSIKA simulated events, even for the great distances that the muons travel. For muons with  momenta > 500 GeV/c the opening angle is less that $1^{0}$ .  A muon with an opening angle of  $1^{0}$ with respect to the cosmic ray primary trajectory whose zenith angle was  $90^{0}$ would pass with a distance of closest approach, $R_{p}$,  of about 22.5 km.  Lower energy muons at larger opening angles of up to  $5^{0}$ could be created closer to HAWC resulting in similar values of $R_{p}$. Therefore using the geometrical calculations from the previous section with the range of possible muon opening angles determined using CORSIKA we have determined that there is no need to generate air showers with $R_{p,max}$ above 25km while still reliably calculating effective areas for higher energy cosmic ray primaries. 

\paragraph*{Generic Cylindrical ``Toy Model'' simulation}
 A simplified ``toy model'' simulation of a perfect cylindrical detector with surface area and height similar to HAWC  provides guidance in the development of the more complete simulation. The simplified model allows us to determine where in phase space to generate extensive air showers that result in nearly-horizontal muons detectable by HAWC. Generating cosmic ray primaries with shower axes over an area perpendicular to the shower direction with a maximum radius,$R_{p,max}$, potentially comparable to the height of the atmosphere (112.8 km) or beyond might be necessary to account for all muons traversing HAWC at nearly-horizontal trajectories. The geometrical considerations of the previous sections indicate that $R_{p,max}$ is likely constrained to be less than 25km. Muon tagging in the toy model is performed by requiring that the length of the muon trajectory intersecting the cylinder exceeds a minimum requirement such as track length >  50m.  A function describing how  effective area increases with $R_{p,max}$ will be determined from the toy model.  Using this function, a corrected effective area scaled to sufficiently large values of $R_{p,max}$ may be performed. This would avoid generating an enormous number of events in a full simulation that is computationally prohibitive. The perfect cylinder toy model simulation is also used to estimate the muon flux impinging upon HAWC, and hence the effective area for a perfect detector, in a given range of zenith angles. The azimuthal response of the actual HAWC detector with volcanoes present determined from a detailed simulation can then be convoluted with the effective area for a perfect cylindrical detector to obtain the effective area as a function of arrival direction for the real HAWC observatory. This simplified model also provides a verification of the investigations using the detailed detector model. 

\paragraph*{Detailed detector simulation} 
HAWCSIM  is  a  GEANT4-based software simulation package internally developed by the HAWC collaboration.  HAWCSIM fully models the instrumental response of HAWC to particles from EAS from CORSIKA simulations as well as the response to single particles.  The estimate of the azimuthal and zenith angle resolution provided by the HAWCSIM simulation studies with single muons  determine the appropriate size binning in arrival directions for an analysis to determine flux as a function of the volcano overburden depth in a given direction. The azimuthal dependence of HAWC's response to nearly-horizontal single muons as determined using HAWCSIM shown in Figure 1b indicates pronounced enhancements in response at the six preferred directions corresponding to an alignment of tanks.  A  response function for arrival directions determined with HAWCSIM response to single muons will be convoluted with the effective area for a perfect cylindrical detector to obtain the effective area as a function of arrival direction for HAWC including detector instrumental effects. For this study, so far, we have only approximately  accounted for the effect of the nearby volcanoes  by considering the material overburden depth in the following manner. The overburden depth as a function of arrival direction has been determined from a digital elevation model \cite{RockDepData} for the region. The resulting minimum muon energy required to penetrate each depth has been determined using the muon energy loss formula from the Particle Data Group \cite{refPDG} and is shown as a function of arrival direction in Figure 2a.   The muon flux reduction factor due to overburden depth  as a function of azimuthal angle for elevation angles between $-1^{0}$ to $5^{0}$  is shown in Figure 2b.  This factor was obtained for a given arrival direction depth by obtaining the expected muon energy distribution from a billion CORSIKA generated protons at large zenith angles with a spectral index of -2.7  after being attenuated by a survival probability of $(1-exp(E_{\mu}/E_{min,depth})$. The factor for a given arrival direction is the ratio of the number of events in the momentum distribution for a given energy cutoff (corresponding to overburden depth) to the number of events in the distribution with an energy cutoff of 50 GeV. The factors for a given azimuthal direction were combined over the chosen range of zenith angles. We will combine the effective area of a perfect cylindrical detector for the detection of nearly-horizontal air showers from cosmic ray primary particles with zenith angles above $85^{0}$ with the azimuthal instrumental response function as determined by HAWCSIM and the azimuthally dependent attenuation due to the volcanoes to determine an approximate effective area as a function of arrival direction for the real HAWC detector with volcanoes. This combined effective area calculation will be used as a cross-check on the result we obtain from the full detector plus volcanoes simulation.  We will also use this simplified simulation chain assuming a standard energy spectrum and composition to predict the number of muons with zenith angles in a restricted range as a function of azimuth angle.  A comparison of this simulated distribution to the observed distribution can be used as an indication of problems with the simulation including hadronic models used in CORSIKA.

We have started the development of a detailed model using GEANT to be incorporated into the HAWCSIM package to better determine the effect of the volcanoes on the response of HAWC to nearly-horizontal muons. The detailed simulation of volcanoes will enable HAWC to be used in a manner similar to an underground detector for nearly-horizontal muon trajectories. We have also developed a modified version of HAWCSIM for this study to rotate the trajectories and momenta of air shower particles at the CORSIKA observation level to generate showers to angles above $90^{0}$  in zenith. We will mirror the thrown events within a few degrees of zenith = $90^{0}$ to account for the muons produced by the cosmic ray primaries slightly beyond the horizon. The use of this full detector plus volcanoes simulation  will provide a  better estimate of HAWC's effective area as a function of arrival direction. The observed muon rate for each energy cutoff, determined by arrival direction, when properly weighted by corresponding effective area and solid angle will be used to measure the energy spectrum of cosmic ray primaries.

\section{Summary}
We have developed simulation tools required to determine the effective area as a function of arrival direction for the detection of cosmic ray primaries that produce nearly horizontal muons.  We are using a toy model simulation to optimize the chosen phase space to process  CORSIKA generated EAS in a full GEANT4-based simulation of the HAWC detector and nearby volcanoes. The toy model simulation also provides a determination of the effective area of a perfect cylindrical detector of HAWC dimensions to cosmic ray primaries that produce nearly-horizontal muons. The effect of the volcanoes on the observed muon rate as a function of arrival direction has been accounted for by considering the effect of overburden depth on required minimum muon energies. Development of the techniques to extract physically meaningful energy spectra and composition measurements is at an early stage but appears promising. 
%%
%%%%%%%%%%%%%%%%%%%%%%%%%%%%%%%%%%%%%%%%%%%%%%%%%%%%%%%%%%
%%
\section*{Acknowledgments}
\footnotesize{
We acknowledge the support from: the US National Science Foundation (NSF); the US Department of Energy Office of High-Energy Physics; the Laboratory Directed Research and Development (LDRD) program of Los Alamos National Laboratory; Consejo Nacional de Ciencia y Tecnolog\'{\i}a (CONACyT), M{\'e}xico (grants 271051, 232656, 260378, 179588, 239762, 254964, 271737, 258865, 243290, 132197), Laboratorio Nacional HAWC de rayos gamma; L'OREAL Fellowship for Women in Science 2014; Red HAWC, M{\'e}xico; DGAPA-UNAM (grants RG100414, IN111315, IN111716-3, IA102715, 109916, IA102917); VIEP-BUAP; PIFI 2012, 2013, PROFOCIE 2014, 2015;the University of Wisconsin Alumni Research Foundation; the Institute of Geophysics, Planetary Physics, and Signatures at Los Alamos National Laboratory; Polish Science Centre grant DEC-2014/13/B/ST9/945; Coordinaci{\'o}n de la Investigaci{\'o}n Cient\'{\i}fica de la Universidad Michoacana. Thanks to Luciano D\'{\i}az and Eduardo Murrieta for technical support.
}

%%
%%%%%%%%%%%%%%%%%%%%%%%%%%%%%%%%%%%%%%%%%%%%%%%%%%%%%%%%%%
%%

\begin{figure}[htbp] 
 \centering
\includegraphics[width=0.49\textwidth]{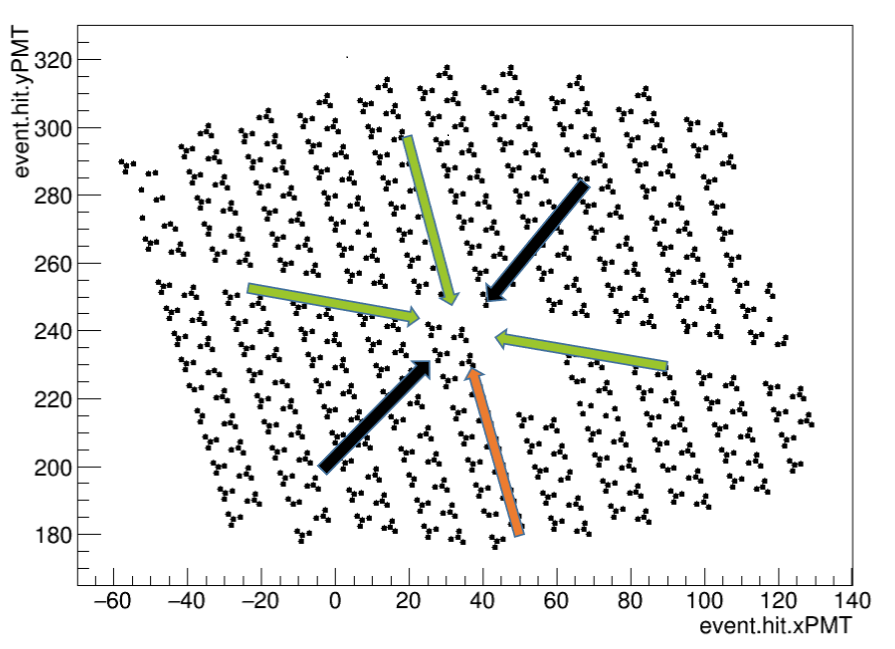} 
  \includegraphics[width=0.49\textwidth]{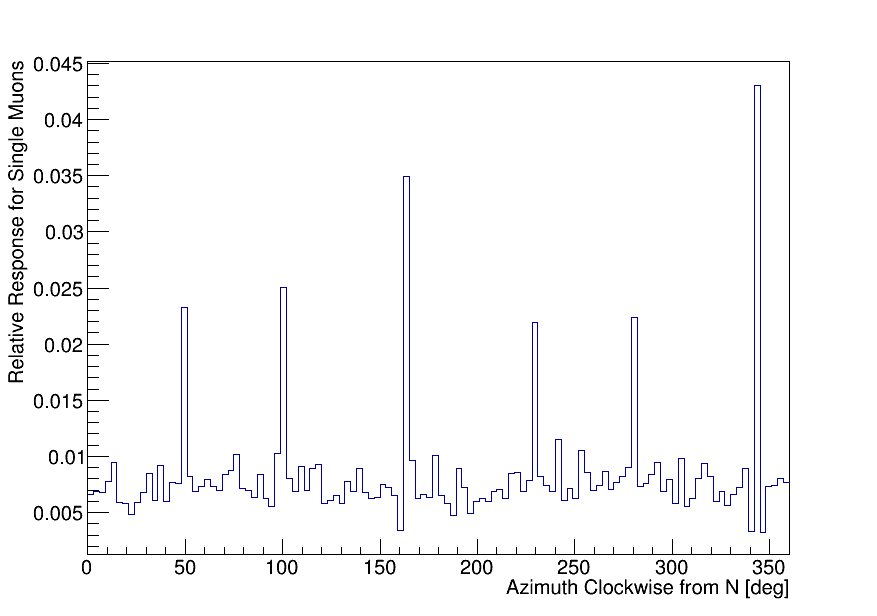} 
  \caption{\textbf{Left Panel a:} Location of the PMTs in the layout of the HAWC water tanks. Each set of 4 PMTs view an isolated volume of water. The azimuthal directions of enhanced response to horizontal muons are indicated by the arrows. \textbf{Right panel b:}   Relative response of identification algorithm  to nearly-horizontal HAWCSIM simulated single muons as a function of azimuth angle. Note the enhanced response in the six directions corresponding to the six preferred directions of tank alignment. }
   \label{fig:Overburden}
\end{figure}

\begin{figure}[htbp] 
 \includegraphics[width=0.49\textwidth]{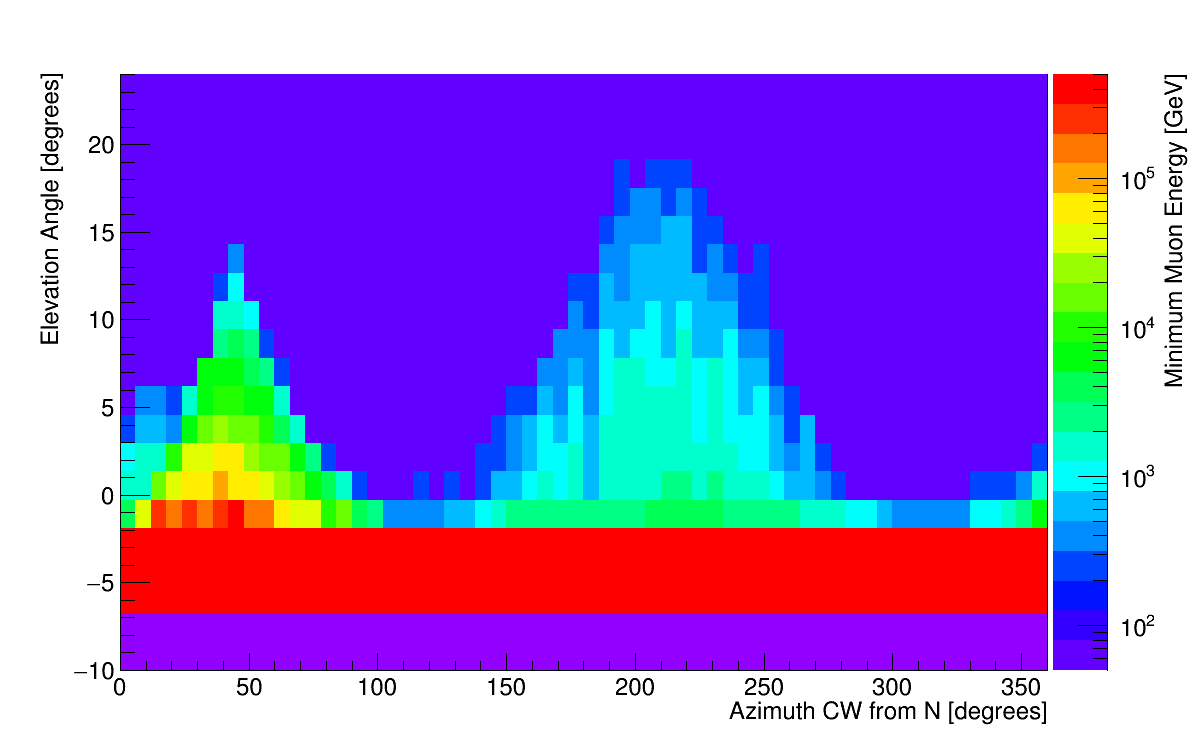} 
  \includegraphics[width=0.49\textwidth]{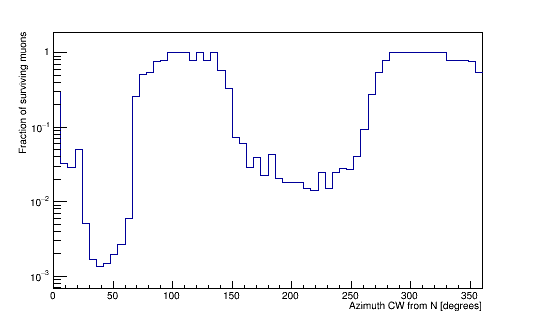}
   % requires the graphicx package
   \caption{\textbf{Left Panel a:} The minimum required muon energy to traverse overburden depth with an exit energy of at least 50 GeV as a function of arrival direction (azimuth CW from N,elevation) due to the volcanoes Pico de Orizaba and Sierra Negra. \textbf{Right Panel b:}   The muon flux reduction factor due to overburden depth  as a function of azimuthal angle for elevation angles between $-1^{0}$ to $5^{0}$ .  This factor was obtained for a given arrival direction depth by obtaining the expected muon energy distribution obtained from a billion CORSIKA generated protons at large zenith angles with a spectral index of -2.7  after being attenuated by a survival probability of $(1-exp(E_{\mu}/E_{min,depth})$. The factor for a given arrival direction is the ratio of the number of events in the momentum distribution for a given energy cutoff (corresponding to overburden depth) to the number of events in the distribution with an energy cutoff of 50 GeV. The factors for a given azimuthal direction were combined over the chosen range of zenith angles.  }
   \label{fig:hawcsim-directions}
\end{figure} 
\end{document}